# A finite element study of the influence of the osteotomy surface on the backward displacement during exophthalmia reduction

Vincent Luboz[1], Annaig Pedrono[2], Dominique Ambard[2], Frank Boutault[3], Pascal Swider[2], Yohan Payan[1]

[1] TIMC-GMCAO Laboratory, UMR CNRS 5525, Faculté de Médecine Domaine de la Merci, 38706 La Tronche, France
{Vincent.Luboz, Yohan.Payan}@imag.fr
http://www-timc.imag.fr/Vincent.Luboz/

[2] Biomechanics Laboratory, IFR30, Purpan University Hospital, 31059 Toulouse, France
pascal.swider@toulouse.inserm.fr

[3] Maxillofacial Department, Purpan University Hospital, 31059 Toulouse, France

**Abstract.** Exophthalmia is characterized by a protrusion of the eyeball. The most frequent surgery consists in an osteotomy of the orbit walls to increase the orbital volume and to retrieve a normal eye position. Only a few clinical observations have estimated the relationship between the eyeball backward displacement and the decompressed fat tissue volume. This paper presents a method to determine the relationship between the eyeball backward displacement and the osteotomy surface made by the surgeon, in order to improve exophthalmia reduction planning. A poroelastic finite element model involving morphology, material properties of orbital components, and surgical gesture is proposed to perform this study on 12 patients. As a result, the osteotomy surface seems to have a non-linear influence on the backward displacement. Moreover, the FE model permits to give a first estimation of an average law linking those two parameters. This law may be helpful in a surgical planning framework.

## 1 Introduction

Exophtalmia is an orbital pathology that affects the ocular muscles and/or the orbital fat tissues [1]. It is characterized by a forward displacement of the eye ball outside the orbit (Fig. 1 (a)). This displacement, called protrusion, may leads to aesthetical problems and to physiological disorders such as a tension of the optic nerve (dangerous for the patient vision) and the ocular muscles and/or an abnormal cornea exposition to the light. Exophthalmia can be due to four causes [1]: a trauma, a tumor, an infection or a disthyroidy. In our works, we mainly focus on the disthyroidian exophthalmia which is the result of an endocrinal dysfunction, as the Basedow illness. This pathology mostly leads to a bilateral exophthalmia since it induces a volume increase of the ocular muscles and/or of the fat tissues.

The classical treatment of exophthalmia is characterized by two steps. The first one aims to stabilize the endocrinal activities (by radiotherapy or surgery). The second step consists in a surgical reduction of the protrusion. The most efficient surgical technique is a decompression of the orbit [2] [3], that is to say an osteotomy of a part of the orbital bone walls via an eyelid incision. It leads to an increase of the orbital volume and thus offers more space to the soft tissues, particularly into the sinuses. To improve the backward displacement of the eye ball, some surgeons push on it in order to evacuate more of the fat tissues in the sinuses (Fig. 1 (b)). The whole procedure is critical since it can produce perturbations in visual functions (e.g. transient diplopia) and may affect important structures such as the optic nerve.

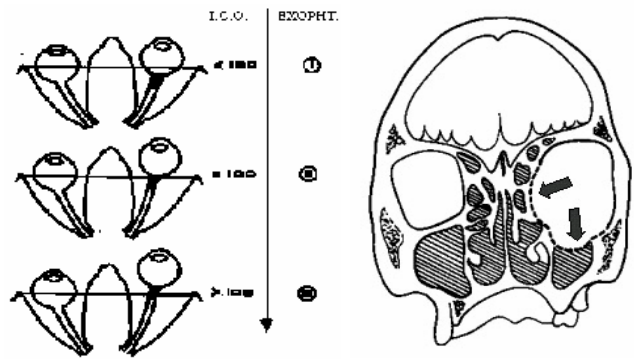

**Fig. 1.** (a) At the left; exophthalmia can be light (top), moderated (middle) or excessive (bottom). (b) At the right; the decompression is performed in the sinuses. The fat tissues then occupied those new cavities

Up to now, the prediction of the results of an exophthalmia reduction is based on clinical observations [4] that gave the following average law: for a 1 $cm^3$ soft tissues decompression, a backward displacement from 1 mm to 1.5 mm is expected. Besides, few works dealing with the biomechanical modeling of the orbital soft tissues have been presented in the literature. [5] and [6] have developed a mass-spring model of the ocular muscles and the eye ball while [7] have proposed an elastic Finite Element (FE) model of the whole orbital soft tissues. Nevertheless, none of these models are applied to the exophthalmia reduction.

In order to improve the exophthalmia reduction planning, a first study has been made by the authors [8] to predict the results of the orbital decompression. In this work, two complementary models have been presented to simulate the surgery. An analytical model that gave a good estimation of the volume decompressed in term of the eye ball backward displacement and a FE model that allowed to simulate the osteotomy and to compute the resulting backward displacement. In this paper, the FE mesh is used to study the influence of the osteotomy surface on the eye ball backward displacement.

## 2 Material and Methods

In the previous work, the FE model was defined to correspond to a specific patient in order to estimate the variations between the simulation results and the clinical measurements, estimated on a pre-operative and a post-operative CT scans. To be as close as possible to the clinical set up for this patient, the morphology, the boundary conditions and the FE parameters were chosen to fit the data measured on this patient and taken as a reference.

The first step was consequently to build the finite element mesh corresponding to this patient. To do this, the surface of the orbit was segmented to get the orbital bones surrounding the soft tissues. This segmentation has been done manually through the definition of a spline on each CT slice since those bones are thin and difficult to automatically separate from the rest of the tissues. The spline set is then extrapolated to give the three-dimensional surface of the bones. The geometry ant the volume of the muscles and the nerve can also be extrapolated with this process.

From this 3D bone surface, the FE mesh can be generated. Since this geometry is complex, no software can be used to automatically mesh it. This process has been done manually using three-dimensional 20-node hexahedrons (quadratic elements). The resulting mesh (Fig. 2 (a) and (b)) is composed of 6948 nodes and 1375 elements.

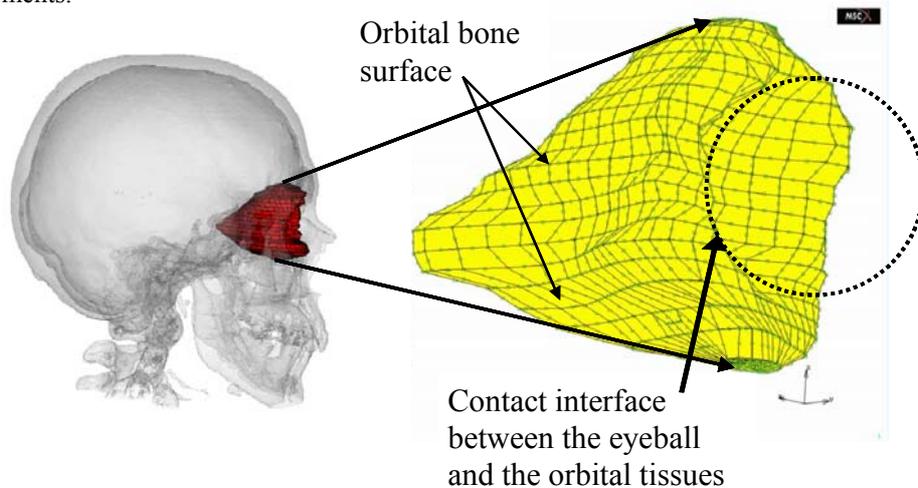

**Fig. 2.** Manual generation of the orbital mesh, (a) left: in the patient skull and (b) right: with a dashed representation of the eyeball

In order to simplify this model, the eyeball was considered as a rigid entity and was thus neglected. Since fat tissues occupy approximately 3/4 of the orbital soft tissue volume and since the fat tissues are predominant in the flow through the osteotomy, a homogenized material modeling the fat tissues has been chosen to simulate the orbital content. Clinical observations [1] describe the orbital fat tissues during a disthyroidy as the combination of an elastic phase composed of fat fibers (mainly collagen) and a fluid phase composed of fat nodules saturated by physiological fluid. Consequently, a poroelastic material [9] has been used to model the intra-orbital soft tissues, using the FE software MARC© (MSC Software Inc.).

The finite element properties of the model has been inspired by the literature ([10], [11] for the Young modulus and [12] for the Poisson ratio) and then adapted to fit the data measured on the pre-operative and post-operative CT scans of the patient studied in the previous work [8]. This adaptation process led to a Young modulus value of 20kPa and a Poisson ratio of 0.1. The two main poroelastic parameters, i.e. the porosity and the permeability, control the fluid behavior through the elastic phase and respectively take into account the fluid retention and the fluid pressure variation. Considering the work of [13] for the inter vertebral disk, the orbital fat tissues permeability was set to 300mm$^4$/N.s and the porosity to 0.4.

Boundary conditions (Fig. 3) have been introduced to simulate the surgery:
The constraint of the bone walls surrounding the soft tissues is translated by a nil displacement and a total sealing effect at the surface nodes.

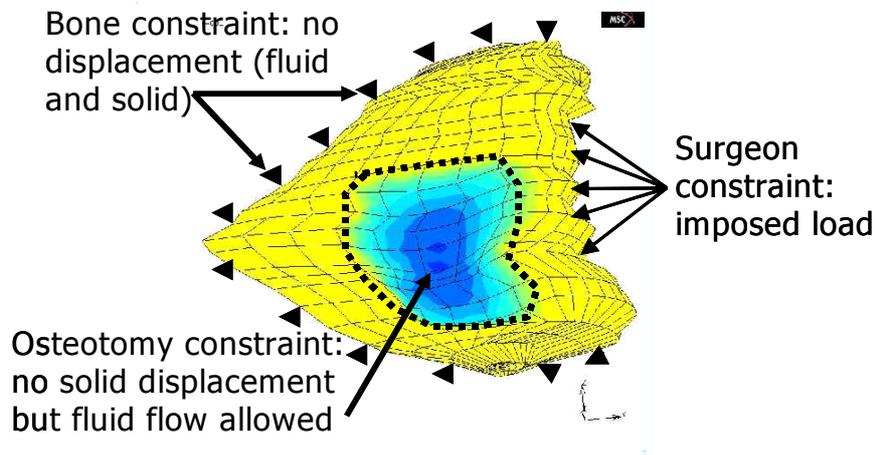

**Fig. 3.** Boundary conditions applied to the orbital mesh to simulate the decompression surgery

Since the periost (the membrane around the soft tissues) remains after the wall osteotomy, it is still constraining the fat tissue elastic phase. Nevertheless, the fluid phase is able to flow through this opening. Consequently, the osteotomy surface nodes are fixed in displacement while they are released in term of sealing effects. To study the osteotomy surface influence on this patient, four different osteotomies have been designed, in accordance with clinical feasibility.

As the soft tissue volume increases, the pressure of these tissues increases too. This overpressure has been estimated at 10kPa by [14]. This value was therefore applied as an initial pore pressure to all the FE mesh nodes. A relaxation time of 2s was then needed to reach pressure equilibrium. During this time, nodes located at the soft tissue/globe interface were free to move.

To simulate the force exerted on the eyeball by the surgeon, an imposed axial load has been applied to all the nodes located at the soft tissue/globe interface (via an external controlled node to simply control this constraint). This imposed load was estimated clinically with a stiffness homemade sensor and gave an average maximum value of 12N. The load constraint is applied in 2s according to a linear ramp from 0 to 12N. Since the eyeball is not modeled no contact analyze is required.

The imposed load is maintained 3s to reach an equilibrium in the tissues and then it is released. To simulate the fact that the fat tissues stay in the sinuses after the decompression, the osteotomy surface was set impermeable to avoid the fluid flowing back into the orbit.

During this simulation, the eyeball backward displacement is estimated through the computation of the displacement of the external node where the imposed load was applied.

This first work gave results close to the data measurements made for the patient studied particularly in terms of backward displacement estimation. Moreover, it appeared that the surface of the osteotomy had a preponderant influence on the eyeball displacement: not surprisingly, a larger osteotomy led to a greater backward displacement.

Nevertheless, since this previous study has been made on only one patient, those results have to be confirmed. This is the aim of this paper which proposes a series of simulations on 12 patients with four different osteotomy surfaces. Fig. 4 shows those osteotomies. Their surfaces have a value of $0.8cm^2$, $1.7cm^2$, $3.4cm^2$ and $5.9cm^2$.

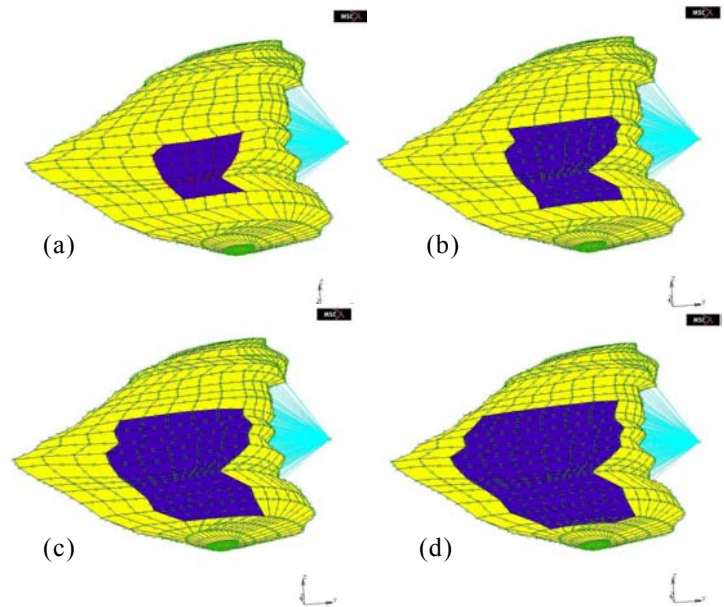

**Fig. 4.** The four osteotomies simulated to study the influence of the surface on the eyeball backward displacement

Since the first patient mesh already exists, an automatic meshing method has been used to generate the 11 new patients meshes (Fig. 5). This method is called the Mesh-Matching algorithm and is described in [15]. This algorithm is based on an optimization method that computes the elastic transformation between a reference mesh and a set of surface points. In this case, the reference mesh is the patient mesh that already exists while the set of surface points is given by the orbit segmentation for the 11 other patients. By using this algorithm, the 11 new FE patient meshes are automatically generate in few minutes enabling to perform the FE simulations of the orbital decompression. The geometry of these meshes is various since the morphology of these patients are different. For example, the volume of the orbital soft tissues ranges from $18.1cm^3$ to $31.4cm^3$.

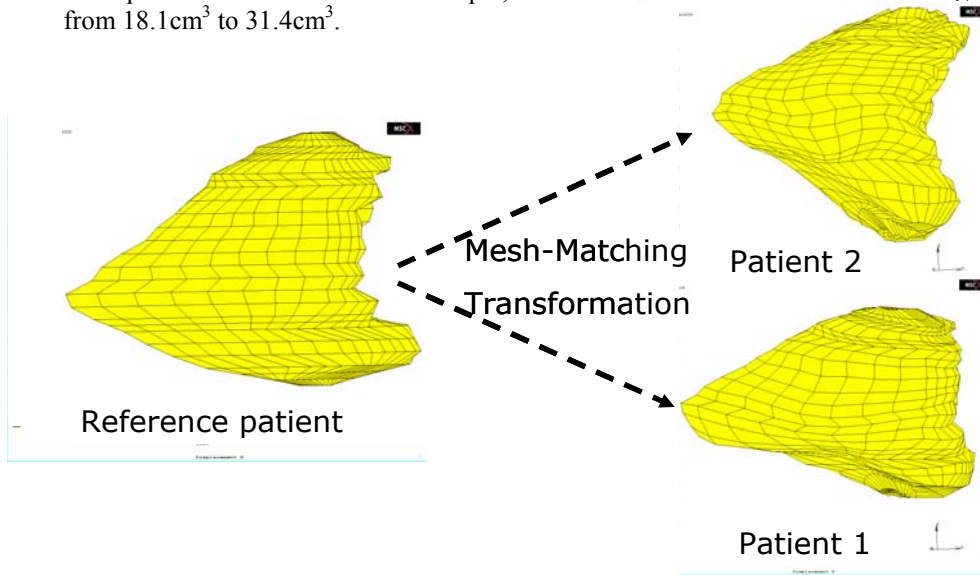

**Fig. 5.** Transformation with the Mesh-Matching of the reference patient mesh in order to automatically generate the new patient meshes and to fit their morphologies. Here, two new patients with various morphologies are generated

Since no post-operative CT scan was performed on these new patients, there was no possibility to analyze the behavior of the soft tissues during and after the surgery. Consequently, the FE parameters can not be evaluated for each of these patients. Though a variation of these parameters may exist clinically, the values used for the first patient were kept for these new patient models.

## 3 Results

Each of these simulations take roughly 1h on a PC equipped with a 1.7GHz processor and 1Go of memory. To estimate the influence of the osteotomy surface on the eyeball backward displacement, a graph showing the relationship between those two variables has been computed (Fig. 6). This graph clearly shows the non linear influence of the surface on the backward displacement. A consequent increase of the osteotomy surface is needed to get a moderate increase of the backward displacement. For example, an osteotomy surface 2 times larger leads to an increase of the eye ball backward displacement of 40%.

Moreover, this graph points out the influence of the patient morphology (the volume, radius and length of the orbit) on the backward displacement. Indeed, the curves presented in Fig. 6 are roughly ordered by the orbital volume, with the smallest volume at the bottom of the graph and the largest orbital volume at the top. A difference of about 50% can be measured between the two extreme patients. Nevertheless, no generic law taking into account the patient morphology influence has been found to match the results of these simulations.

Consequently, an average law has been computed to estimate the relationship between the osteotomy surface and the eyeball backward displacement. This law is the equation of the average curve computed from the 12 patient curves. It is not a precise estimation of the clinical behaviour of the orbital soft tissues during the surgery since it is far from the extreme values as shown on the Fig. 6. This equation gives the backward displacement *disp* in function of the osteotomy surface *surf*:

$$disp = 1.1 * \ln(surf) + 1.9 \ . \tag{1}$$

The following equation seems more useful to a surgeon that the 1cm$^3$ versus 1/1.5mm relation proposed by [4]. Indeed, it gives an estimation of the osteotomy surface, *surf*, to do in order to obtain a suited eyeball backward displacement, *disp*:

$$surf = e^{\frac{disp - 1.9}{1.1}} \ . \tag{2}$$

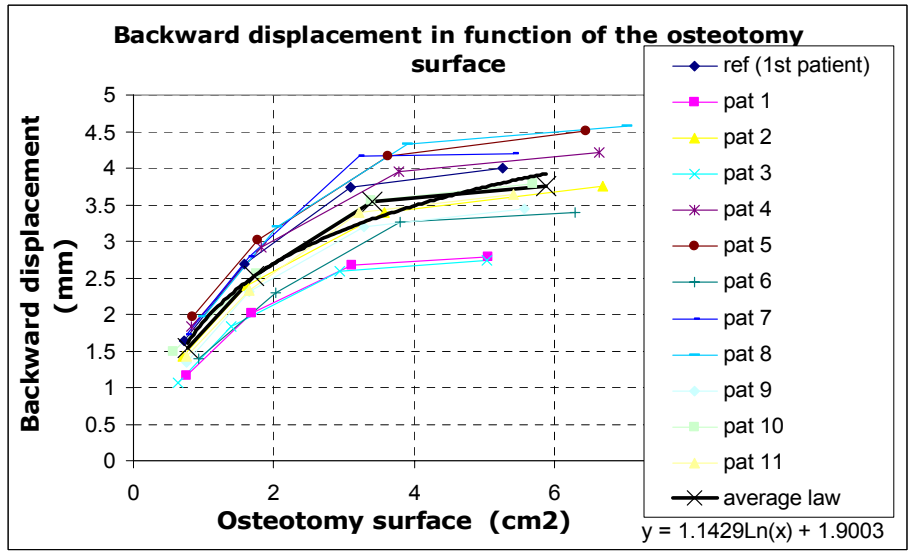

**Fig. 6.** Influence of the osteotomy surface on the eyeball backward displacement for the 12 patients and average curve

## 4 Discussion

The FE model presented in this paper seems to be able to answer the main question raised by the surgeons during the exophthalmia decompression: what osteotomy surface has to be done to get a suited backward displacement? The answer is estimated with equation (2). Indeed, the previous study has shown that the results given by the poroelastic model were close to the clinical measurements. Even if no measurements have been done for the 11 new patients, because of the absence of post-operative CT scans, it can be assumed that the results plotted in the Fig. 6 are not too far from the real soft tissues behavior.

With this FE model, the non linear influence of the osteotomy surface seems non neglectable. Indeed, an important increase of this surface must be done to get an effect on the eyeball backward displacement. Given that this surface cannot be extended infinitely, since the orbital cavity is physiologically small, it could be interesting for a surgeon to know the relationship between the osteotomy surface and the backward displacement during the surgery planning step. The equation (2) gives that relationship in real time and therefore could help a surgeon to take into account the influence of the surface to choose and, if it is necessary, to increase the osteotomy size knowing its relative influence on the eyeball displacement.

Nevertheless, it has been shown that equations (1) and (2) do not take in consideration the influence of the patient morphology. Those equations are consequently imprecise. To be more precise, this study has to be done on more patients and with comparisons with post-operative CT scans, to check the validity of the model results. Those new simulations would permit to take into account the influence of the orbital patient morphology and consequently to improve the estimation of the two equations presented above.

Another important assumption has been made during the definition of the patient models. Indeed, the same FE parameters for the orbital soft tissues have been used for each patient. Clinically this assumption does not seem to be valid since the orbital soft tissues tension and the degree of the muscle fibrosis are heterogeneous and may vary from one patient to another. Here again, a most important study, with post-operative scans, would be able to take into account this fact and to determine the inter-patient FE parameter variations.

## 5 Conclusion

This paper has proposed an estimation of the influence of the osteotomy surface on the eyeball backward displacement in the context of orbital decompression. This evaluation is based on a poroelastic finite element model of the orbital soft tissues, presented in a previous study. It seems to point out that the relationship between the surface and the backward displacement is non linear (an important increase of the surface leads to a moderate increase of the displacement) and that the patient morphology may also influence the eyeball displacement. The equation (2) gives a first real time estimation of the law linking the backward displacement to the osteotomy surface. This equation may be helpful for a surgery planning and may replace the actual law stating that for a 1 cm$^3$ soft tissues decompression, a backward displacement from 1 mm to 1.5 mm is expected. Nevertheless, the study will have to be performed on a more important patient set to verify and to improve this estimation. The Mesh-Matching algorithm will consequently be useful in the future since it gives an easy process to generate automatically new patient meshes.

Future works will thus focus on a further study on a more important set of patients with post-operative scans to (i) try to determine the influence of the patient morphology on the backward displacement (and improve equation (2)), (ii) estimate the variations of the FE parameters amongst different patients and (iii) validate the results of this study.

As a perspective, the integration of the eyeball and the muscles in the FE mesh will be studied to take into account their actions on the soft tissues behavior. A contact fluid/structure model may consequently be developed and compared to the poroelastic model.